\documentclass[10pt,final,journal,letter,oneside,twocolumn]{IEEEtran}

\usepackage[noadjust]{cite}
\usepackage{srcltx}
\usepackage{psfrag}
\usepackage{epsfig}
\usepackage[tbtags,sumlimits,nointlimits,reqno]{amsmath}
\usepackage{amssymb}
\usepackage{xcolor}
\usepackage{float}
\usepackage{subfigure}
\usepackage{soul}
\usepackage{footnote}
\usepackage{amsmath}

\setstcolor{red}

\usepackage{showkeys}   

\begin{document}

\title{Generalized Coupled-line All-Pass Phasers}

\author{%
       Shulabh~Gupta,~\IEEEmembership{Member,~IEEE,} Qingfeng Zhang,~\IEEEmembership{Member,~IEEE,} Lianfeng~Zou,~\IEEEmembership{Student Member,~IEEE,} Li Jun~Jiang,~\IEEEmembership{Senior Member,~IEEE,} and~Christophe~Caloz,~\IEEEmembership{Fellow,~IEEE}
\thanks{S. Gupta and L. J. Jiang are with the Electrical and Electronic Engineering Department of The University of Hong Kong, China. Email: shulabh@hku.hk.}
\thanks{Q. Zhang is with the Department of Electrical and Electronics Engineering, South University of Science and Technology of China (SUSTC), Shenzhen, China.}
\thanks{L. Zou, and C. Caloz are with the Department of Electrical
Engineering, Poly-Grames Research Center, \'{E}cole Polytechnique de Montr\'{e}al, Montr\'{e}al, QC, Canada.}
}
\markboth{IEEE Transactions for Microwave Theory and Techniques~2013}{Shell \MakeLowercase{\textit{et al.}}: Bare Demo of IEEEtran.cls for
Journals}

\maketitle

\begin{abstract}
Generalized coupled-line all-pass phasers, based on transversally-cascaded (TC), longitudinally-cascaded (LC) and hybrid-cascaded (HC) coupled transmission line sections, are presented and demonstrated using analytical, full-wave and experimental results. It is shown that for $N$ commensurate coupled-line sections, LC and TC phasers exhibit $N$ group delay peaks per coupled-line section harmonic frequency band, in contrast to the TC configuration, which exhibits only one peak within this band. It is also shown that for a given maximum achievable coupling-coefficient, the HC configuration provides the largest group delay swing. A wave-interference analysis is finally applied to the various coupled-line phasers, explaining their unique group delay characteristics based on physical wave-propagation mechanisms.
\end{abstract}


%
\IEEEpeerreviewmaketitle
%

\section{Introduction}

Radio-Analog Signal Processing (R-ASP) has recently emerged as a new paradigm for monitoring, manipulating and processing radio signals in real time~\cite{Caloz_MM_2012,Caloz_PIEEE_10_2011,Lewis_SAW_OSP_2005}. Compared to conventional Digital Signal Processing (DSP) techniques, R-ASP operates on signals directly in their pristine analog form to execute specific operations enabling microwave or millimeter-wave and terahertz applications. It thus provides a potential solution, specially at high frequencies, to overcome the drawbacks of DSP techniques, which include high-cost A/D and D/A conversion, high power consumption, low-speed and high complexity.

The heart of an R-ASP system is a phaser, which is a component exhibiting a specified frequency-dependent group-delay response within a given frequency range~\cite{Caloz_MM_2012}. When a broadband signal propagates through a phaser, its spectral components progressively separate from one another in the time domain due to their different group velocities, a well-known phenomenon commonly referred to as chirping. This spectral discrimination in the time domain allows manipulating spectral bands individually and directly in the time domain, thereby enabling several high-speed and broad-band processing operations. Some recently demonstrated R-ASP applications include signal receivers and transmitters for communications~\cite{Abielmona_TMTT_11_2009,Nguyen_MWCL_08_2008}, frequency meters and discriminators for cognitive networks~\cite{Nikfal_TMTT_06_2011,Nikfal_MWCL_11_2012}, spectrum analyzers for instrumentation~\cite{Gupta_TMTT_04_2009,Schwartz_MWCL_04_2006,Laso_TMTT_03_2003}, signal manipulation for efficient signal processing~\cite{Xiang_TMTT_11_2012,Schwartz_MWCL_01_2008} and chip less tags for radio-frequency identification (RFID) systems~\cite{Gupta_AWPL_11_2011}.

Phasers can be either of reflective type or transmission type. Reflective-type phasers are single port structures which are converted into two-port structures using a broadband circulator or a hybrid coupler. They are mostly based on the principle of Bragg reflections, and include microstrip chirped delay lines~\cite{Laso_MWCL_12_2001}, artificial dielectric substrate based phasers~\cite{Coulombe_TMTT_08_2009} and reflection-type waveguide phasers~\cite{Zhang_TMTT_08_2012}. While reflection-type phasers impose less constraints on the design parameters of the phaser compared to transmission-type phasers~\cite{Zhang_EL_14_2013}, their requirement of an external one-port to two-port conversion component, incurring additional loss along with undesired phase distortions in the overall delay response, is a major drawback. Transmission-type phasers, on the other hand, are inherently two-port components. Surface acoustic wave (SAW) devices~\cite{Campbell_SAW_1989}, and magneto-static devices~\cite{Ishak_PIEE_02_1988} are some classical representatives of these phasers, but they are suitable only for very low frequencies and narrow-bandwidth applications. While some recently proposed transmission-type phasers, based on coupled-matrix analysis, offer great synthesis flexibility for advanced group delay engineering, they are usually restricted to narrow-band designs~\cite{Zhang_TMTT_3_2013,Atia_TMTT_08_2002}. For broader-band applications, coupled-line all-pass phasers are more suitable, offering also greater design simplicity and benefiting from efficient synthesis procedures~\cite{Cristal_TMTT_06_1966,Gupta_TMTT_09_2010,Horii_MWCL_01_2012,Gupta_IJCTA_2013}.

The coupled-line all-pass phasers reported to date are based on \emph{transversal cascaded (TC)} C-sections and/or D-sections synthesizing prescribed group delay responses~\cite{Gupta_IJCTA_2013,Zhang_IJRMCAE_2013}. On the other hand, a \emph{longitudinal cascade (LC)} of commensurate coupled transmission-lines was first demonstrated in \cite{Cristal_TMTT_06_1966}. In this work, a third type of cascading configuration is proposed based on a combination of a TC and an LC configuration, hereby termed as \emph{Hybrid cascade (TC)}. The corresponding coupled-line phaser is called a Hybrid-cascaded (HC) coupled-line phaser. The HC coupled-line phaser was first introduced in~\cite{Gupta_Folded_ICEAA_2013} and later used in group delay engineering in~\cite{Paradis_EL_00_2013}. This set of three cascading configurations, LC, TC and HC, represent the fundamental cascading schemes upon which more complex phasers can be constructed, leading to vast variety of coupled-line all-pass phasers with rich and exotic dispersion characteristics. Further, the group delay characteristics of these three cascaded configurations are investigated and compared in details and their unique properties are explained using a rigorous wave-interference analysis.

The paper is organized as follows. Section~II introduces the TC, LC and HC coupled-line phaser topologies, with a comparison of their group delay characteristics based on their analytical transfer functions. It also presents corresponding fabricated prototypes along with measured results, validating the analytical models and confirming the various group delay characteristics. Section~III presents the wave-interference mechanisms that may be used to construct a general transfer function with coupled-line all-pass phasers. These mechanisms are then used to explain some unique group delay features of LC and HC coupled-line phasers. Finally, Sec.~IV provides conclusions.

\section{Coupled-Line All-Pass Phasers}

\subsection{Basic Transfer Functions}

All-pass transfer functions are based on two building-block transfer functions, the C-section and the D-section transfer functions~\cite{Steenart_TMTT_01_1963}. 

\begin{figure}[htbp]
\begin{center}
\psfrag{a}[c][c][0.8]{$\ell=\lambda_g/4$@$\omega_0$}
\psfrag{b}[l][c][0.8]{$k_1$}
\psfrag{c}[l][c][0.8]{$k_2$}
\psfrag{x}[c][c][0.8]{$v_\text{in}$}
\psfrag{y}[c][c][0.8]{$v_\text{out}$}
\includegraphics[width=\columnwidth]{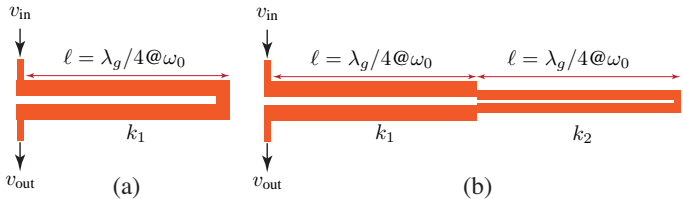}
\caption{Transmission-type all-pass phaser topologies. a)~C-section and b)~D-section, with $\omega_0$ being the quarter-wavelength frequency of a transmission line.} \label{Fig:CDSection}
\end{center}
\end{figure}

A C-section transfer function can be realized by a two-port transmission-line structure, consisting of a (four-port) coupled-line coupler with its through and isolated ports interconnected by an ideal transmission-line section, as shown in Fig.~\ref{Fig:CDSection}(a). Its transfer function can be derived by applying the interconnection boundary condition between the ingoing and outgoing waves at the two end ports of the coupler. The C-section transfer function is given by

\begin{subequations}
\begin{equation}
S_{21}(\theta) = \left(\frac{\rho - j\tan\theta}{\rho + j\tan\theta}\right),\label{Eq:Csection}
\end{equation}
\begin{equation}
\text{with }\rho= \sqrt{\frac{1+k}{1-k}},
\end{equation}
\end{subequations}

\noindent where $k$ is the coupling coefficient between the coupled lines forming the structure. Its all-pass nature, $|S_{21}|=1,\,\forall\theta$, is immediately verified by noting that the magnitudes of the numerator and denominator in~\eqref{Eq:Csection} are equal, and it may also be verified that the function provides a frequency-dependent group delay response reaching a maximum value at \mbox{$\theta=m\pi/2$}, where $m$ is an integer. This condition corresponds to frequencies where the length of the coupled-line section is an odd multiple of a quarter wavelength. Upon the high-pass to low-pass transformation $s = j\tan\theta$, the C-section is seen to be a first-order phaser, with one real pole and one real zero, placed symmetrically about the imaginary axis in the $s-$plane.

The derivation of a generalized transfer function corresponding to a coupled-line coupler terminated with an arbitrary all-pass load with transfer function $S_0$ is provided in App.~\ref{App:CSec_TF}, and reads

\begin{equation}
S_{21}(\theta) = b+\frac{a^2S_0}{1-bS_0},\label{Eq:S21_S0}
\end{equation}

\noindent where in~\eqref{Eq:S21_S0} $a$ and $b$ are the through and coupled transfer functions of the coupler without the end connection. The C-section transfer function is a particular case of this network with $S_0=1$.

The second-order transmission-line all-pass phaser is the D-section, which consists of a coupled-line coupler terminated with a C-section, as shown in Fig.~\ref{Fig:CDSection}(b). Using the general transfer function form \eqref{Eq:S21_S0} with $S_0$ given by \eqref{Eq:Csection}, the transfer function of a D-section is found as

\begin{subequations}
\begin{equation}
S_{21}(\theta) = \left(\frac{1 - \rho_a\tan^2\theta - j\rho_b\tan\theta}{1 - \rho_a\tan^2\theta + j\rho_b\tan\theta}\right),\label{Eq:Dsection}
\end{equation}

with

\begin{equation}
\rho_a= \sqrt{\frac{1-k_1}{1+k_2}}\sqrt{\frac{1+k_1}{1-k_2}}
\end{equation}
 
and

\begin{equation}
\rho_b= \sqrt{\frac{1-k_1}{1+k_1}} + \sqrt{\frac{1-k_2}{1+k_2}}
\end{equation}\label{Eq:rhoab}

\end{subequations}

\noindent where $k_{1,2}$ are the coupling coefficients of the two sections. Upon the transformation $s = j\tan\theta$, the D-section is seen to be a second-order phaser, with two pairs of complex para-conjugate zeros and poles in the $s-$plane.

\subsection{Phaser Topologies}

\begin{figure}[htbp]
\begin{center}
\subfigure[]{
\psfrag{a}[l][c][0.8]{$\ell_1$, $k_1$}
\psfrag{b}[l][c][0.8]{$\ell_2$, $k_2$}
\psfrag{c}[l][c][0.8]{$\ell_3$, $k_3$}
\psfrag{d}[l][c][0.8]{$\ell_n$, $k_N$}
\psfrag{e}[l][c][0.8]{$\ell_4$, $k_4$}
\psfrag{f}[l][c][0.8]{$\ell_{N-1}$, $k_{N-1}$}
\psfrag{x}[c][c][0.8]{$v_\text{in}$}
\psfrag{y}[c][c][0.8]{$v_\text{out}$}
\psfrag{m}[c][c][0.8]{\shortstack{Transversally-cascaded\\ configuration}}
\psfrag{n}[c][c][0.8]{ \shortstack{Longitudinally-cascaded\\ configuration}}
\psfrag{p}[c][c][0.8]{\shortstack{Hybrid-cascaded \\configuration}}
\includegraphics[width=\columnwidth]{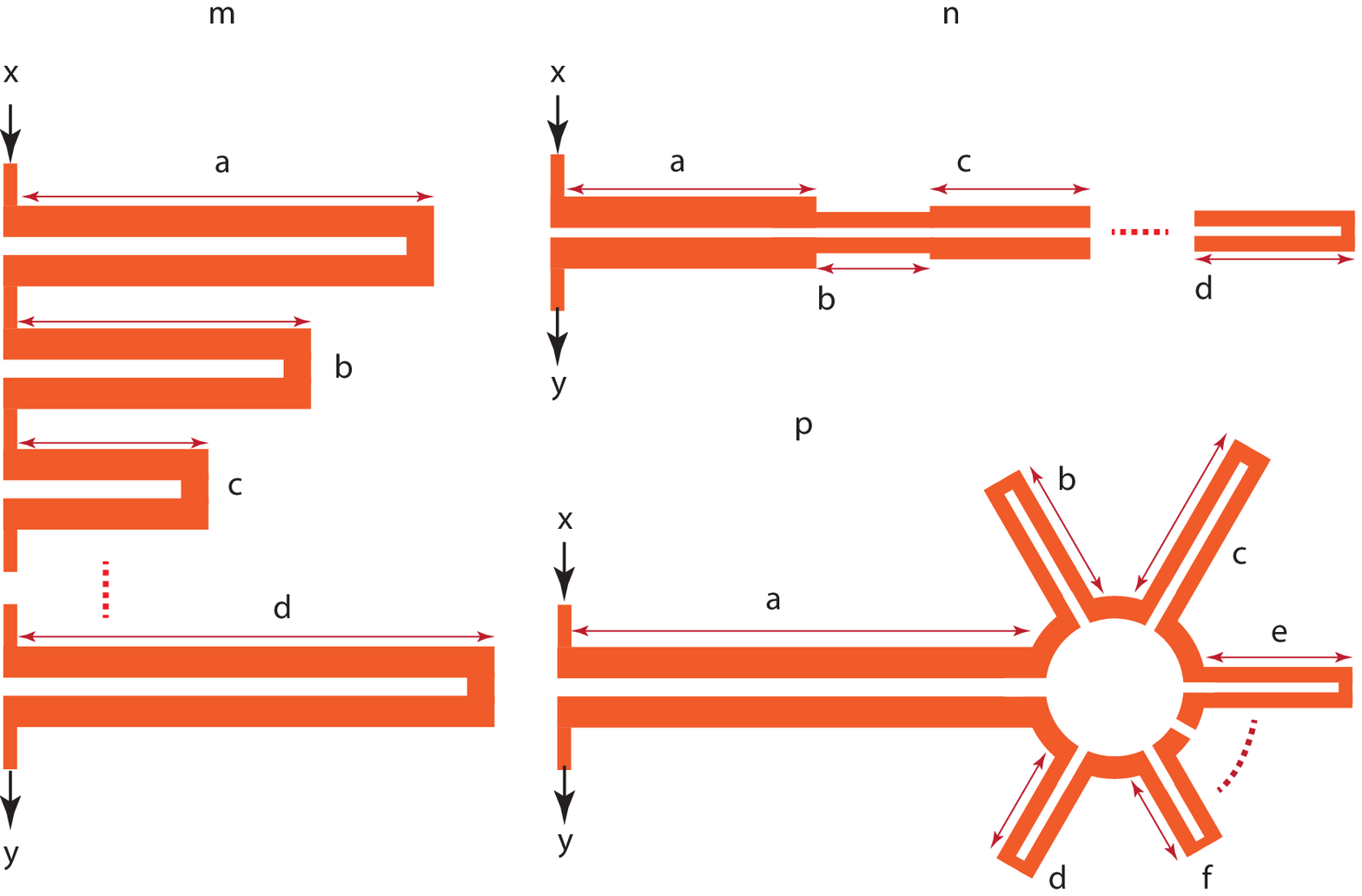}}
\subfigure[]{
\psfrag{a}[c][c][1]{Electrical length $\theta\times\pi$ (rad)}
\psfrag{b}[c][c][1]{Group delay $\times d\theta/d\omega$}
\psfrag{c}[l][c][1]{\shortstack{$k_1 = 0.7$\\$k_2=0.6$\\$k_3=0.5$\\$k_4=0.4$}}
\psfrag{d}[l][c][0.8]{TC coupled-lines}
\psfrag{e}[l][c][0.8]{LC coupled-lines}
\psfrag{f}[l][c][0.8]{HC coupled-lines}
\psfrag{g}[c][c][0.8]{$\ell_1  = \ell_2 =\ell_3 = \ell_4 = \ell$}
\includegraphics[width=0.8\columnwidth]{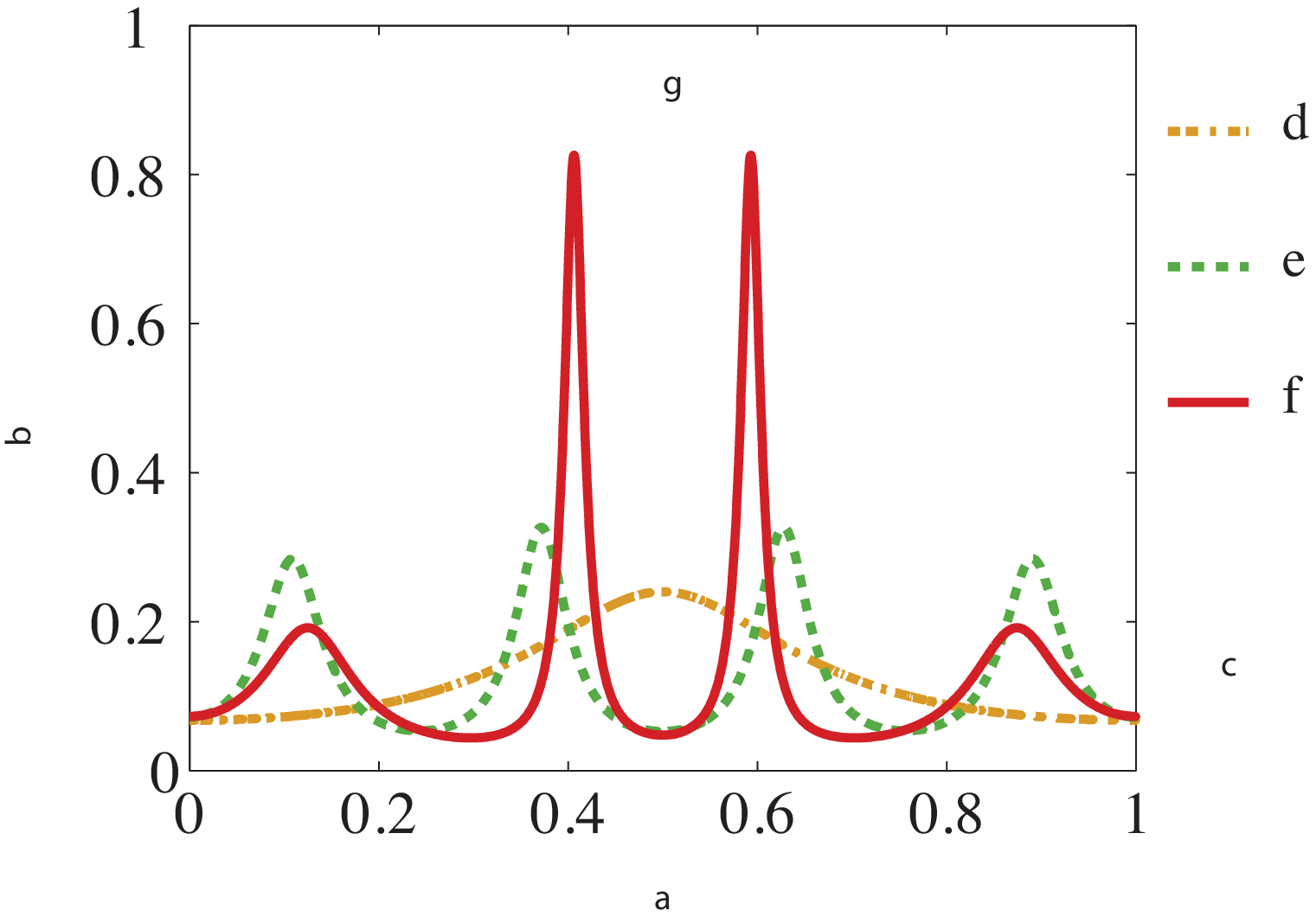}}
\caption{Cascaded coupled-line all-pass phasers. a) Generic topologies for a transversally-cascaded (TC), longitudinally-cascaded (LC) and hybrid-cascaded (HC) coupled-line phasers. b) Typical group delay response of the three phaser topologies in (a) over the lowest coupled-line section harmonic frequency band with $N=4$ coupled-line sections of identical length $\ell_i  = \ell$.} \label{Fig:Phasers}
\end{center}
\end{figure}

The group delay profiles of a C-section and a D-section have a specific and restricted shape that depend on the length and coupling coefficients of the sections involved. However, combining several coupled-line sections in an appropriate fashion allows synthesizing virtually arbitrary prescribed group delay responses between the input and the output ports within a given frequency range. Such phasers can be conceptualized and categorized based on how the different coupled-line sections are connected together. The three basic configurations shown in Fig.~\ref{Fig:Phasers} are possible, and are next described.

\begin{enumerate}

\item \emph{Transversally-cascaded (TC) coupled-line phaser}. In this configuration several C-sections are cascaded in the direction that is transverse to the axis of the transmission lines forming the C-sections, as shown in Fig.~\ref{Fig:Phasers}(a). The corresponding transfer function is then simply the product of the individual C-section transfer functions and thus reads 

\begin{equation}
S_{21}^\text{TC}(\theta) = \prod_{i=1}^N \left(\frac{\rho_i - j\tan\theta}{\rho_i + j\tan\theta}\right),\label{Eq:TF_TC}
\end{equation}

\noindent where $\rho_i = \sqrt{(1+k_i)/(1-k_i)}$, $k_i$ being the coupling coefficient of the $i^\text{th}$ section.

\item \emph{Longitudinally-cascaded (LC) coupled-line phaser}. In this configuration, several coupled-line sections are cascaded in the direction of the transmission lines forming the coupled-lines, as shown in Fig.~\ref{Fig:Phasers}(a), with the last section being a C-section \cite{Cristal_TMTT_01_1969}. The corresponding transfer function can be derived by iteratively constructing the load transfer functions starting from the last coupled-line section towards to the input port, as derived in App.~\ref{Sec:Series_TF}. The result is

\begin{equation}
S_{21}^\text{LC}(\theta) = S_1(\theta) = b_1 + \frac{a_1^2S_2(\theta)}{1- b_1S_2(\theta)},\label{Eq:Series_TF}
\end{equation}

\noindent where $S_2(\theta)$ is the overall transfer function of the structure starting from the $2^\text{nd}$ to the $N^\text{th}$ coupled-line section. It is to be noted that a D-section is the particular case of an LC coupled-line phaser with $N=2$.

\item \emph{Hybrid-cascaded (HC) coupled-line phaser}. This configurations consists of a combination of TC and LC coupled-line sections, as illustrated in Fig.~\ref{Fig:Phasers}(c). This may be seen as a coupled-line coupler terminated with a TC coupled-line phaser. The corresponding transfer function can be written using the TC coupled-line transfer function as $S_0$ in (\ref{Eq:S21_S0}), leading to

\begin{subequations}
\begin{equation}
S_{21}^\text{HC}(\theta) = b_1 + \frac{a_1^2S_0(\theta)}{1- b_1S_0(\theta)},
\end{equation}
\begin{equation}
S_0(\theta) = \prod_{i=2}^N \left(\frac{\rho_i - j\tan\theta}{\rho_i + j\tan\theta}\right).
\end{equation}
\end{subequations}

\end{enumerate}

The diversity of possible transfer functions provided by the TC, LC and HC coupled-line phasers for general parameters is too great to be tractable in such a paper. Therefore, for the sake of simplicity and comparability, we shall restrict our analysis to the case of commensurate coupled-line sections (i.e. coupled-line sections having all the same length). However, it should be kept in mind that the above analytical transfer functions are general and that the corresponding phasers exhibit much richer synthesis possibilities than those presented next.

\subsection{Group Delay Response}

The three configurations in Fig.~\ref{Fig:Phasers}(a) exhibit very different group delay responses, as shown in Fig.~\ref{Fig:Phasers}(b) for the case of $N=4$ and $\ell_1=\ell_2=\ell_3=\ell_4=\ell$. Under the latter condition, the TC coupled-line phaser has a delay shape similar to that of a regular C-section, with a single delay maximum over the lowest coupled-line section harmonic frequency band, $\theta\in[0,\pi]$. In contrast, the LC configuration exhibits 4 delay peaks, within the same bandwidth. These peaks are quasi-uniformly spaced and quasi-equal in magnitude, with group delay swings, $\Delta\tau = \tau_\text{max} - \tau_\text{min}$, that are larger than those obtained in the TC case\footnote{Note that the response can be drastically different in the case of non-commensurate sections.}. Similar to the LC case, the HC coupled-line phaser exhibits 4 delay peaks, but with dramatically different characteristics. While the outermost delay peaks are lower, the center ones have a steep slope within a narrow bandwidth and thus exhibit a very large delay swing $\Delta \tau$. Moreover, it is observed that the two central peaks are closer to the each other compared to the LC case. Thus, for a given maximum achievable coupling $k$, the HC coupled-line phaser provides the largest group delay swing $\Delta\tau$, among the three configurations.

\begin{figure}[htbp]
\begin{center}
\psfrag{a}[c][c][1]{Electrical length $\theta\times\pi$ (rad)}
\psfrag{b}[c][c][1]{Group delay $\times d\theta/d\omega$}
\psfrag{c}[c][c][0.8]{$N=7$}
\psfrag{d}[c][c][0.8]{$N=11$}
\psfrag{e}[c][c][0.8]{$N=15$}
\psfrag{f}[c][c][0.6]{(TC coupled-lines)}
\psfrag{g}[c][c][0.6]{(LC coupled-lines)}
\psfrag{h}[c][c][0.6]{(HC coupled-lines)}
\includegraphics[width=\columnwidth]{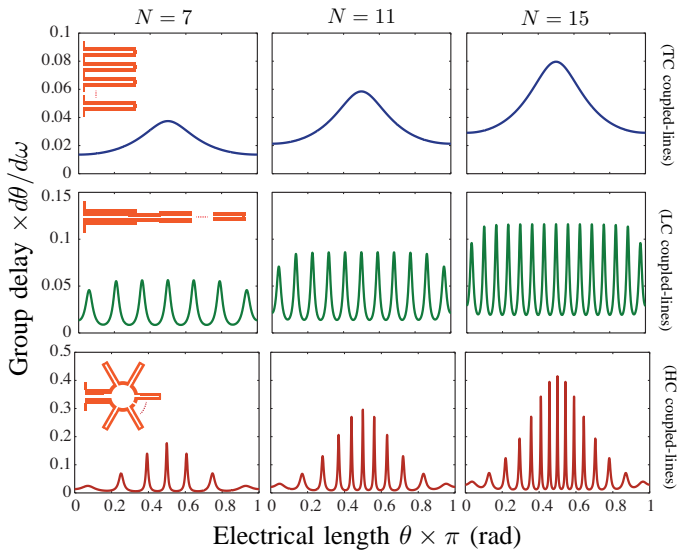}
\caption{Typical group delay responses for the different cascaded coupled-line phasers in Fig.~\ref{Fig:Phasers}(a) with different numbers $N$ of coupled-line sections of length. Here all the phaser sections have the same length $\ell$ while $k_1 > k_2 > k_3 \ldots > k_N$.} \label{Fig:PhaserParametric}
\end{center}
\end{figure}

Figure~\ref{Fig:PhaserParametric} shows the typical group delay responses of these three configurations for different number of coupled-line sections $N$. While the TC coupled-line phaser maintains a single delay peak across the bandwidth for all $N$'s, the number of delay peaks in the LC and the HC case scales with $N$. In general, for $N$ coupled-line sections, the group delay exhibits $N$ delay peaks within a periodic band. The trends observed in Fig.~\ref{Fig:Phasers}(b) are confirmed:

\begin{enumerate}
\item While the delay values of the peaks in the LC case are near identical, except for the $1^\text{st}$ and the $N^\text{the}$ peak, a strong variation of the delay peak values occurs in the HC case, with the central peak exhibiting the largest group delay swing.
\item The HC coupled-line phaser provides the largest group delay swing, and the largest absolute delay values, among the three cases, at the cost of locally reduced bandwidth around the peak values.
\item Compared to the LC coupled-line phaser, the spacing between adjacent peaks across the bandwidth is strongly non-uniform in the case of HC coupled-line phasers, with more crowding near the centre of the periodic bandwidth.
\end{enumerate}

Despite the very different group delay characteristics, the following theorem is common to the three phaser configurations: \emph{Given $N$ sections, the total area under the $\tau-\theta$ curve is $N\pi$ regardless of the configuration}. This theorem is demonstrated in App.~\ref{Sec:ConstArea}. As a result, based on the aforementioned considerations regarding the number of delay peaks, compared to the weakly-dispersive TC coupled-line phasers, LC and the HC phasers make a more efficient use of their delay-frequency area by shaping the delay curve so as to produce higher dispersion regions. In other words, they locally enhances the group delay peaks around specific frequencies by reducing the other peaks, so as to keep the total area constant. A detailed wave-interference mechanism will be used in Sec.~III to explain the other delay characteristics of these phaser configurations.

\begin{figure*}[htbp]
\begin{center}
\psfrag{a}[c][c][0.8]{$\varepsilon_r = 4.38$}
\psfrag{b}[l][c][0.6]{Rogers 4003C}
\psfrag{h}[c][c][0.8]{$h$}
\psfrag{c}[l][c][0.6]{Epoxy layer}
\psfrag{d}[l][c][0.6]{GND trace}
\psfrag{e}[l][c][0.6]{17~$\mu$m Cu trace}
\includegraphics[width=2\columnwidth]{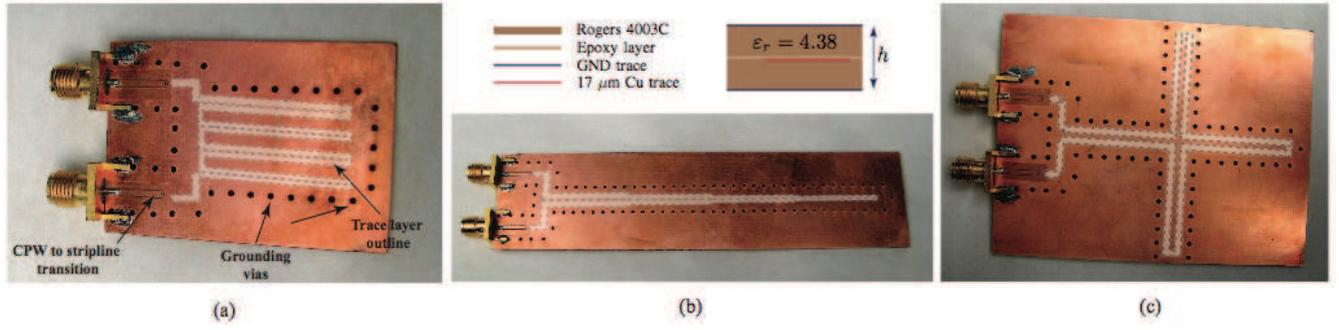}
\caption{Photographs of coupled-line phaser prototypes in stripline technology consisting of $N=4$ coupled-line sections in three different configurations. a) TC coupled-line phaser. b) LC coupled-line phaser, and c) HC coupled-line phaser. The pictures show the grounding vias and outline of the copper traces sandwiched between the two substrate layers. Linewidth and line-gap of each coupled-line section are: $\ell_i\in[22\;22\;22\;22]$~mils and $g_i\in[22\;17\;12\;7]$~mils, which corresponds to the coupling coefficients $k_i\in[0.065\;0.095\;0.145\;0.215]$ extracted from full-wave simulation (FEM-HFSS). Length of each section is 1000~mils. (a)~TC phaser. (b)~LC phaser. (c)~HC phaser.} \label{Fig:FabPics}
\end{center}
\end{figure*}

\subsection{Experimental Illustrations}

In order to confirm the delay characteristics of the three coupled-line configurations, their prototype were built, in stripline technology, as shown in Fig.~\ref{Fig:FabPics}. The prototypes consist of two RO4003C substrate layers, each 20~mils thick. The stripline is fed through a coplanar-waveguide to stripline transition at each input and output port. An array of conducting vias following the signal layer profile, were used to maintain the same potential between the top and ground planes.

\begin{figure}[htbp]
\begin{center}
\psfrag{a}[c][c][0.8]{group delay $\tau$ (0.2 ns/div)}
\psfrag{b}[c][c][0.8]{frequency (GHz)}
\psfrag{c}[l][c][0.7]{TC coupled-lines}
\psfrag{d}[l][c][0.7]{LC coupled-lines}
\psfrag{e}[l][c][0.7]{HC coupled-lines}
\psfrag{f}[l][c][2.8]{$\}$}
\psfrag{g}[l][c][0.7]{Measured}
\psfrag{h}[l][c][0.7]{FEM-HFSS}
\psfrag{i}[l][c][0.7]{Theory}
\psfrag{j}[l][c][0.8]{$S_{21}$ (dB)}
\psfrag{k}[l][c][0.8]{$S_{11}$ (dB)}
\includegraphics[width=\columnwidth]{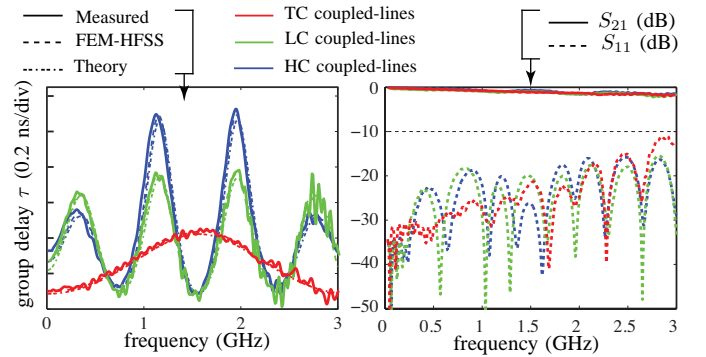}
\caption{Measured S-parameters and group delay responses of the prototypes shown in Fig.~\ref{Fig:FabPics} compared with full-wave and theoretical results.} \label{Fig:FabN4}
\end{center}
\end{figure}

\begin{figure}[htbp]
\begin{center}
\subfigure[]{
\psfrag{a}[c][c][0.8]{\color{white}$N=1$}
\psfrag{b}[c][c][0.8]{\color{white}$N=2$}
\psfrag{c}[c][c][0.8]{\color{white}$N=3$}
\psfrag{d}[c][c][0.8]{\color{white}$N=4$}
\psfrag{e}[c][c][0.8]{\color{white}$N=5$}
\psfrag{f}[c][c][0.8]{\color{white}$N=6$}
\psfrag{g}[c][c][0.8]{\color{white}$N=7$}
\psfrag{h}[c][c][0.8]{\color{white}$N=8$}
\psfrag{i}[c][c][0.8]{\color{white}$N=9$}
\psfrag{x}[l][c][0.8]{\color{white}Port 1}
\psfrag{y}[l][c][0.8]{\color{white}Port 2}
\includegraphics[width=0.75\columnwidth]{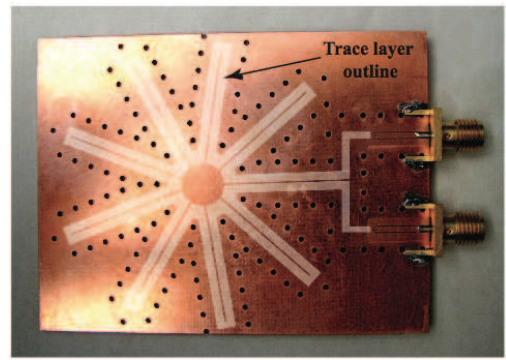}}
\subfigure[]{
\psfrag{a}[c][c][0.8]{group delay $\tau$ (1 ns/div)}
\psfrag{b}[c][c][0.8]{frequency (GHz)}
\psfrag{c}[l][c][0.7]{Meas.}
\psfrag{d}[l][c][0.7]{FEM-HFSS}
\psfrag{e}[l][c][0.7]{Theory}
\psfrag{f}[c][c][0.8]{S-parameters (dB)}
\psfrag{g}[l][c][0.8]{$S_{21}$}
\psfrag{h}[l][c][0.8]{$S_{11}$}
\includegraphics[width=\columnwidth]{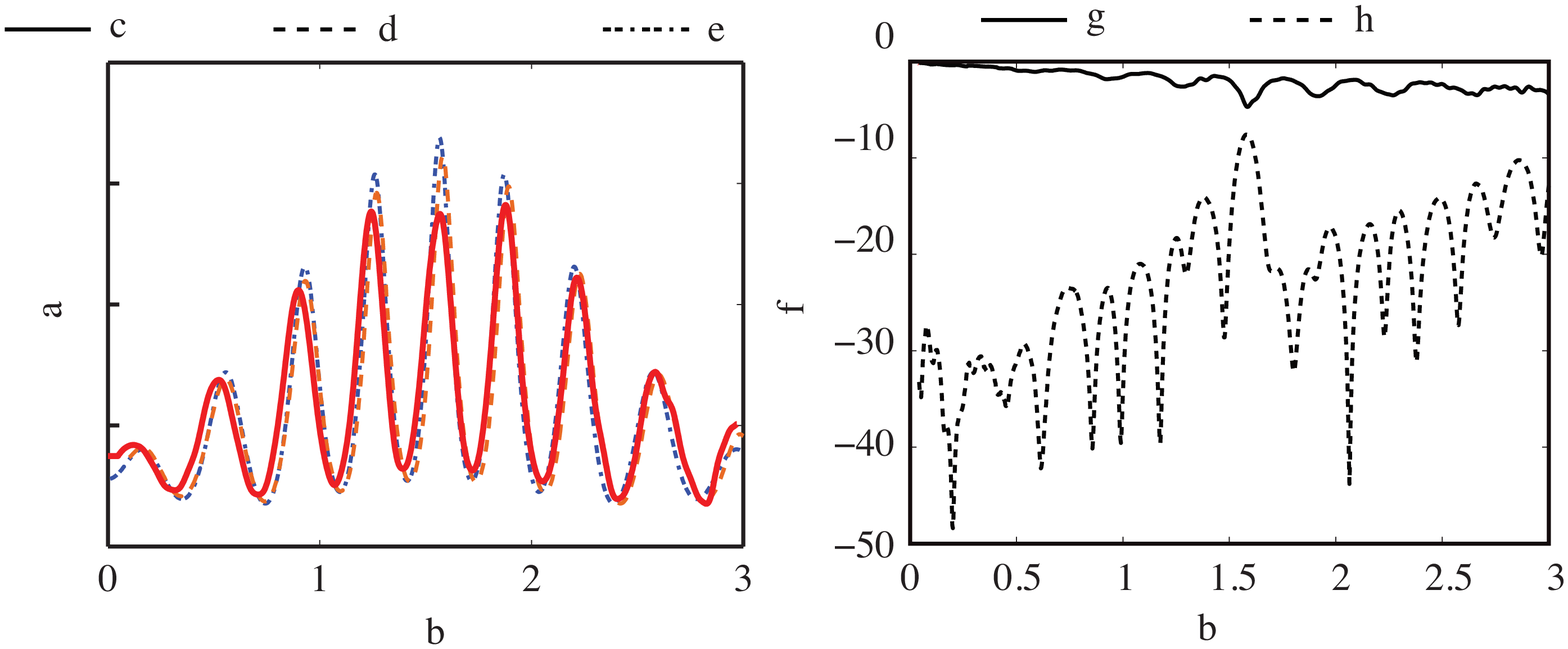}}
\caption{HC coupled-line phaser with $N=9$ coupled-line sections. a) Photographs. b) Measured group delay response. c) Measured S-parameters. Linewidth, line-gap and length of every section are 20~mils, 8~mils and 1000~mils, respectively.} \label{Fig:FabN9}
\end{center}
\end{figure}

Figure~\ref{Fig:FabN4} shows the measured S-parameters and the group delay responses for the three prototypes of  Fig.~\ref{Fig:FabPics}, corresponding to the TC-, LC- and HC coupled-line phasers for $N=4$. All prototypes are well-matched across the entire bandwidth of interest with $S_{11}<-10$~dB in all cases. An excellent agreement between measured results, full-wave and the analytical results of Sec.~II-B, is observed in all cases. As concluded from Fig.~\ref{Fig:Phasers}(b), HC coupled-line phasers provide the largest group delay swing among all three configurations. Fig.~\ref{Fig:FabN9}(a) shows another prototype for the HC coupled-line phaser with $N=9$. Again, measured S-parameters exhibit reasonable agreement within the design bandwidth, as seen in Fig.~\ref{Fig:FabN9}(b), with excellent agreement between the measured and simulated group delay responses. It should be particularly noted that, despite their simplicity, the analytically derived transfer functions predict the various group delay characteristics in a convincing manner in all cases. These formulas may thus be deemed appropriate for fast and efficient designs of coupled-line phasers.

\section{Wave-interference Explanation}

\subsection{Signal Flow Analysis}

The transfer function of a coupled-line coupler terminated with an all-pass transfer function $S_0$ was derived in App.~\ref{App:CSec_TF} using a scattering matrix approach. An alternative approach to obtain such a transfer function is using the signal flow graph analysis in conjunction with wave interference consideration~\cite{Zhang_APM_2_2013,Gupta_TMTT_12_2012}. The signal flow graph analysis provides deeper insight into the wave propagation mechanisms involved and is thereby instrumental to unveil the group delay characteristics of coupled-line phasers.

\begin{figure}[htbp]
\begin{center}
\psfrag{a}[c][c][1]{$\ell$}
\psfrag{b}[c][c][0.8]{$k$}
\psfrag{c}[c][c][0.8]{$k$}
\psfrag{d}[c][c][0.8]{$k$}
\psfrag{f}[c][c][0.8]{$a$}
\psfrag{g}[c][c][0.8]{$b$}
\psfrag{e}[c][c][0.8]{$S^2$}
\psfrag{x}[c][c][1]{$v_\text{in}$}
\psfrag{y}[c][c][1]{$v_\text{out}$}
\includegraphics[width=\columnwidth]{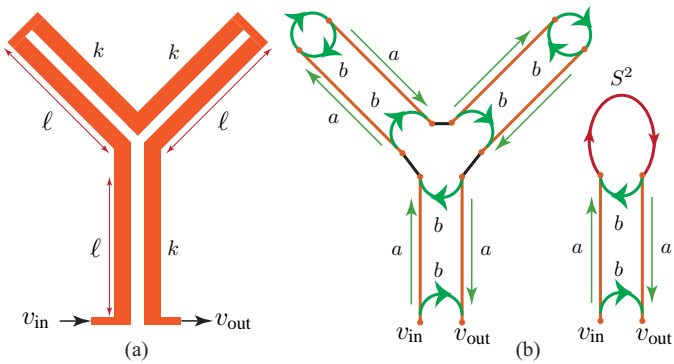}
\caption{Wave interference phenomenology in HC coupled-line phasers. a)~Phaser layout for $N=3$ coupled-line sections of identical lengths and couplings $k$. b)~Corresponding signal flow graph in terms of the coupled-transfer function $b$ and the through transfer function $a$. $S$ is the overall transfer function of a single C-section.} \label{Fig:N3Phaser}
\end{center}
\end{figure}

Let us consider an HC coupled-line phaser with \mbox{$N=3$}, composed of ideally matched coupled-line sections with infinite isolation, as shown in Fig.~\ref{Fig:N3Phaser}(a). Let us also assume that all coupled sections have the same length and coupling coefficient. The structure may be seen as a coupled-line coupler terminated with a pair of TC C-sections. The corresponding signal flow graph is shown in Fig.~\ref{Fig:N3Phaser}(b). In the terminating C-sections, the through signal component is coupled back into the structure via the end connection, resulting in the formation of signal loops, which provides the necessary variation in the delay across the design bandwidth~\cite{Gupta_TMTT_12_2012}. The problem can be simplified by considering a net transfer function $S^2$ representing the complete termination, as indicated in Fig.~\ref{Fig:N3Phaser}(b).

Based on the simplified signal flow of Fig.~\ref{Fig:N3Phaser}(b), the contributions of the different waves along the structure may be summed up to build the overall transfer function $S_{21}$ as

\begin{align}\label{eq:S21_sfg_constr}
S_{21}(\theta) &= \overbrace{b}^\text{direct coupled} + \overbrace{(a\times S^2 \times a)}^\text{direct through}  \nonumber
\\&\qquad + \overbrace{(a\times S^2 \times b \times S^2 \times a)}^\text{first loop} \nonumber
\\&\qquad+ \overbrace{(a \times S^2 \times b \times S^2 \times b \times S^2 \times a)}^\text{second loop} + \ldots\nonumber
\\& = b + a^2S^2[1 + bS^2 + b^2S^4 + b^3S^6 + \ldots] \nonumber\\
&= b + \frac{a^2S^2}{1- bS^2},
\end{align}

\noindent which is identical to~\eqref{Eq:S21_S0} obtained using the scattering matrix method with $S_0=S^2$, where $S_0$ may represent an arbitrary all-pass function. Since the overall transfer function is all-pass, it may be expressed as $e^{j\phi}$ ($|e^{j\phi}|=1,\,\forall\phi$), and~\eqref{eq:S21_sfg_constr} may be compactly rewritten as

\begin{subequations}
\begin{equation}
S_{21}(\theta) = e^{j\phi} = b + a^2S_0 S_\text{loop}\label{Eq:WE_TF_AP}
\end{equation}
with
\begin{equation}
S_\text{loop} = \sum_{n=0}^\infty b^nS_0^{n}.
\end{equation}
\end{subequations}

\noindent Differentiating~\eqref{Eq:WE_TF_AP}, using the definition $\tau(\theta) = -d\phi/d\omega$, and re-arranging the terms, yields

\begin{subequations}
\begin{equation}
\begin{split}
\tau(\theta) = je^{-j\phi}&\left\{\overbrace{\frac{db}{d\theta}}^\text{Term I} + \overbrace{\frac{a^2}{b}\left(S_0\frac{db}{d\theta} + b\frac{dS_0}{d\theta}\right)\Gamma_\text{loop}}^\text{Term II} \right.\\
&\quad\left. + \overbrace{S_\text{loop}\left(2aS_0\frac{da}{d\theta} + a^2\frac{dS_0}{d\theta}\right)}^\text{Term III}\right\},
\end{split}\label{Eq:Tau_N_S}
\end{equation}
with
\begin{equation}
\Gamma_\text{loop} = \sum_{n=0}^\infty nb^nS_0^n.
\end{equation}
\end{subequations}

\noindent In the particular case $S_0=1$, the HC coupled-line phaser reduces to the single C-section with the group delay~\cite{Gupta_TMTT_12_2012}

\begin{equation}
\tau(\theta) = je^{j\phi}\left(\frac{db}{d\theta} + \frac{a^2}{b}\frac{db}{d\theta} \Gamma_\text{loop} + 2a\frac{da}{d\theta} S_\text{loop} \right).\label{Eq:Tau_N_1}
\end{equation}

Comparing the delay expression~\eqref{Eq:Tau_N_S} of HC phasers to the group delay \eqref{Eq:Tau_N_1} of a C-section, it appears that both Term~II and Term~III have an extra factor proportional to $dS_0/d\theta$ contributing to the delay swing when the termination is dispersive, i.e. $dS_0/d\theta\ne 0$. This reveals that the greatness of the group delay swings observed in the LC and HC coupled-line phasers in Sec.~II are essentially due to the dispersive nature of the terminations. In contrast, in the TC coupled-line phaser case, which have regular C-sections with non-dispersive terminations, i.e. $S_0=1$, the coefficients proportional to $dS_0/d\theta$ in Terms~II and III vanish, resulting in smaller delay swings.

\subsection{Wave Interference Phenomenonology in HC and LC Coupled-Line Phasers}

Consider again a coupled-line coupler terminated with load of arbitrary all-pass transfer function and a corresponding signal flow graph of the type shown in~Fig.~\ref{Fig:N3Phaser}(b). Without loss of generality, consider the termination to be $S_0$ instead of $S^2$ to represent a general termination. Depending on $S_0$, the phaser may be either an LC or an HC coupled-line phaser. Following the signal flow graph, the total group delay, $\tau(\theta)$, is given by (\ref{Eq:Tau_N_S}), with $S_\text{loop} = \sum_{0}^\infty b^nS_0^n$ and $\Gamma_\text{loop} = \sum_{0}^\infty nb^nS_0^n$.

Since the load transfer function, $S_0$, is assumed to be all-pass and, in general, dispersive, it may be written as $S_0(\theta) = e^{j\phi_0(\theta)} = e^{-j\int\tau_0d\theta}$. Two extreme cases may then be distinguished:

\begin{enumerate}

\item Case I: if $\phi_0 = 2m\pi$ and $\theta\ne n\pi$ (ensuring $b\neq 0$), where $m$, $n$ are integers, $S_0=+1$, so that
\begin{align}
&S_\text{loop} = 1 + b + b^2 + b^3 +  b^4 + b^5 + b^6 + \ldots \notag \\
&\Gamma_\text{loop} = b + 2b^2 + 3b^3 +  4b^4 + 5b^5 + 6b^6 + \ldots.\notag
\end{align}

\noindent This represents a \emph{constructive} loop-interference situation where both $S_\text{loop}$ and $\Gamma_\text{loop}$ are maximized, resulting in a \emph{maximum group delay} according to~\eqref{Eq:Tau_N_S}.

\item Case II: if $\phi_0 = (2m+1)\pi$ and $\theta\ne n\pi$, where $m$, $n$ are integers, $S_0=-1$, so that
\begin{align}
&S_\text{loop} = 1 - b + b^2 - b^3 +  b^4 - b^5 + b^6 \ldots \notag \\
&\Gamma_\text{loop} = - b + 2b^2 - 3b^3 +  4b^4 - 5b^5 + 6b^6 \dots.\notag
\end{align}

\noindent This represents a \emph{destructive} loop-interference situation where both $S_\text{loop}$ and $\Gamma_\text{loop}$ are minimized, resulting in a \emph{minimum group delay} according to~\eqref{Eq:Tau_N_S}.
\end{enumerate}

\begin{figure}[htbp]
\begin{center}
\psfrag{a}[c][c][0.8]{$\tau[S_{21}]$ (arb. units)}
\psfrag{c}[c][c][0.8]{electrical length $\theta\times\pi$}
\psfrag{b}[c][c][0.8]{$\phi_0(\theta) \times \pi$}
\psfrag{d}[c][c][0.8]{$\tau_0(\theta)$}
\psfrag{x}[c][c][0.8]{$S_0$}
\includegraphics[width=\columnwidth]{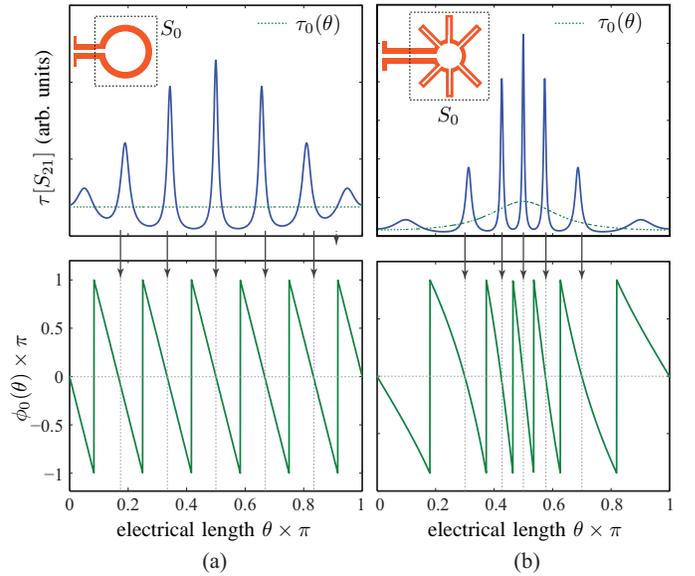}
\caption{Typical group delay response of a coupled-line phaser for the case of a)~a non-dispersive and b)~a dispersive load, $S_0$.} \label{Fig:WE_GD}
\end{center}
\end{figure}

Let us examine the different types of coupled-line phasers at the light of these observations. First, consider the case of a non-dispersive termination, $S_0$ i.e a termination with a constant non-zero group delay. Figure~\ref{Fig:WE_GD}(a) shows the typical delay response $\tau[S_{21}]$ of such a phaser. Two observations can be made:

\begin{itemize}
\item The group delay peaks occur around the regions where the phase of the load $\phi_0$ is a multiple of $2\pi$, as expected from Case~II above, and vice-versa for the group delay minima. This is exactly the case for the highest peaks and progressively less the case for lower peaks.
\item The group delay peaks are uniformly spaced, which is an expected result from the fact that $S_0$ is non-dispersive (linear phase $\phi_0$).
\end{itemize}

Now consider a dispersive termination $S_0$ as in an HC coupled-line phaser. Figure~\ref{Fig:WE_GD}(b) shows the typical delay response in such a case. The dispersive characteristics of $S_0$ manifests itself in the non-constant nature of the delay $\tau_0(\theta)$ and in the strong wavelength compression region around $\theta=\pi/2$. Again, the locations of the various group delay peaks of the phaser occur around the regions where the phase of the termination $\phi_0$ is $2m\pi$, and vice-versa for the group delay minima. Furthermore, the adjacent delay peaks are now non-uniformly spaced, with closely packed peaks in the highly dispersive region of $S_0$. In conclusion, while the locations of $2m\pi$ phase values in $\phi_0$ determine the locations of the delay peaks of the phaser, the dispersive nature of the termination $S_0$ controls the spacing between the adjacent peaks.

\begin{figure}[htbp]
\begin{center}
\psfrag{q}[c][c][0.8]{$S_{21}^\text{LC}$}
\psfrag{r}[c][c][0.8]{$S_0$}
\psfrag{s}[c][c][0.8]{$S_1$}
\psfrag{t}[c][c][0.8]{$S_2$}
\psfrag{u}[c][c][0.8]{$S_3$}
\psfrag{v}[c][c][0.8]{$S_4$}
\psfrag{w}[c][c][0.8]{$S_5$}
\psfrag{a}[c][c][0.8]{electrical length $\theta\times\pi$}
\psfrag{b}[c][c][0.8]{$S_0^c$}
\psfrag{c}[c][c][0.8]{$\tau_0(\theta)$ (arb. units)}
\psfrag{e}[c][c][0.8]{$\tau[S_{21}^\text{LC}](\theta)$ (arb. units)}
\psfrag{d}[c][c][0.8]{$\phi_0(\theta)$ (arb. units)}
\psfrag{x}[c][c][0.6]{C-section}
\psfrag{y}[c][c][0.6]{LC coupled-line}
\includegraphics[width=\columnwidth]{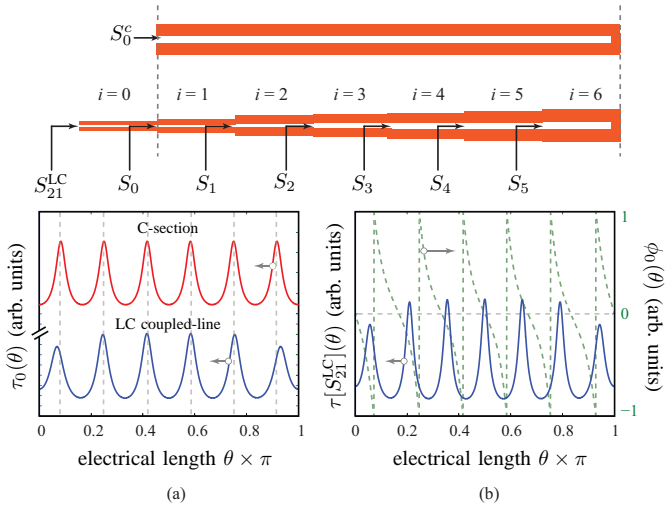}
\caption{Typical group delay response of an LC coupled-line phaser. b) Group delay of $S_0$ in an LC coupled-line phaser, compared to that of a regular C-section of equivalent length. b) Group delay of the LC coupled-line phaser related to the transmission phase of $S_0$. All the lengths are assumed to be equal and $k_i<k_{i+1}\forall i$ in the case of an LC coupled-line phaser.} \label{Fig:WE_GD_LC}
\end{center}
\end{figure}

Finally, consider an LC coupled-line phaser, as illustrated in Fig.~\ref{Fig:WE_GD_LC}. In this case, the transfer function of the load of the $i=0$ coupled-line section is the transfer function of another LC coupled-line phaser formed between the $1^\text{st}$ section and the $6^\text{th}$ section. Compared to the HC coupled-line phaser, the closed form expression of $S_0$ is not readily available, as it must constructed iteratively, as done in App.~\ref{Sec:Series_TF}. In this situation, a more qualitative approach may be followed to deduce the effects of $S_0$ in an LC configuration. The termination function $S_0$ may be seen as a regular C-section, having a transfer function $S_0^C$, of same size with \emph{small perturbations} in the coupling coefficients along the structure. Since, a conventional C-section has a periodic delay response with uniformly spaced peaks, the LC coupled-line phaser must exhibit a similar delay pattern in the small perturbation limit. Figure~\ref{Fig:WE_GD_LC}(a) confirms this prediction, as $\tau_0(\theta)$ closely resembles the delay response of a regular C-section, $\tau_0^C(\theta)$, except at the two extreme ends where the peaks are slightly smaller. Once the termination $S_0$ is known, the delay response of the overall phaser can be easily found. Figure~\ref{Fig:WE_GD_LC}(b) shows the overall group delay of the phaser, and as described above, the locations of the various group delay peaks of the phaser occur around the regions where the phase of the termination $\phi_0$ is a multiple of $2\pi$. Moreover, since these $2m\pi$ locations of $\phi_0(\theta)$ are quasi-uniformly spaced, the resulting delay peaks in $\tau[S_{21}]$ are also quasi-uniformly separated in $\theta$.

The LC coupled-line phaser results in Fig.~\ref{Fig:WE_GD_LC} might give the impression that the LC configuration provides little benefit over a simple C-section, since responses shown are very similar. However, the results of Fig.~\ref{Fig:WE_GD_LC} are obtained for coupled-line sections of identical length and small coupling variations, for the sake of the phenomenological explanation. However, if the section lengths are allowed to be different from each other and if the coupling coefficients are allowed to vary more, the LC configuration may synthesize a great diversity of group delay functions, which is clearly impossible using the single C-section of corresponding length~\cite{Cristal_TMTT_06_1966}.

\section{Conclusion}

Generalized coupled-line all-pass phasers, based on
transversally-cascaded (TC), longitudinally-cascaded (LC) and hybrid-cascaded (HC) coupled transmission line sections, have been presented and demonstrated using analytical, full-wave and experimental results. The corresponding analytical transfer functions have been derived using matrix methods which have been found to accurately model the unique group delay characteristics of these phasers. It has been shown that in contrast to the TC phaser, LC and HC coupled-line phasers, consisting of $N$ coupled-sections exhibit $N$ group delay peaks within a harmonic frequency band. Moreover, for a given maximum achievable coupling-coefficient, the HC configuration provides the largest group delay swing, at the expense of reduced bandwidth, around the peak locations. This follows from the fact that the area under the group delay curve is $N\pi$ regardless the configuration.

\begin{figure}[htbp]
\begin{center}
\psfrag{y}[c][c][0.8]{$v_\text{out}$}
\psfrag{x}[c][c][0.8]{$v_\text{in}$}
\psfrag{a}[c][c][0.8]{T-cascade}
\psfrag{c}[c][c][0.8]{H-cascade}
\psfrag{b}[c][c][0.8]{L- cascade}
\includegraphics[width=0.7\columnwidth]{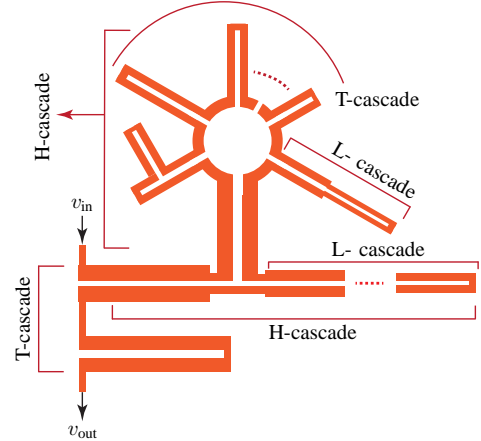}
\caption{Generalized coupled-line all-pass phaser consisting of unlimited combinations of TC-, LC- and HC coupled-line structures.} \label{Fig:Final}
\end{center}
\end{figure}

Based on the typical group delay characteristics, it has been shown that the TC configuration is best suited for broad-band phasers, while the LC and TC configurations are more suitable for narrow-band applications requiring large group delay swings. A rigorous wave-interference analysis has been applied to the coupled-line phasers so as to provide deep insight into the delay mechanisms and the unique group delay characteristics of LC and HC coupled-line phasers, based on wave propagation phenomenology. The TC, LC and HC configurations represent the three fundamental cascading schemes upon which more complex phasers maybe constructed to obtain diverse and rich dispersion characteristics. An illustration of such a construction is shown in Fig.~\ref{Fig:Final}, which combines TC, LC and HC topologies into a general coupled-line phaser configuration. Such a configuration enables virtually unlimited group delay responses, and are anticipated to be useful in synthesizing efficient transmission-line phasers for various R-ASP applications.

\section{Appendix}\label{Sec:Appendix}


\subsection{Derivation of the C-Section Transfer Function [Eq.~(\ref{Eq:Csection})]}\label{App:CSec_TF}

\begin{figure}[htbp]
\begin{center}
\subfigure[]{
\psfrag{y}[c][c][0.8]{$\ell, k$}
\psfrag{x}[c][c][0.8]{$S_0$}
\psfrag{a}[c][c][0.7]{$\psi_1^+$}
\psfrag{b}[c][c][0.7]{$\psi_1^-$}
\psfrag{c}[c][c][0.7]{$\psi_2^+$}
\psfrag{d}[c][c][0.7]{$\psi_1^-$}
\psfrag{e}[c][c][0.7]{$\psi_3^+$}
\psfrag{f}[c][c][0.7]{$\psi_1^-$}
\psfrag{g}[c][c][0.7]{$\psi_4^+$}
\psfrag{h}[c][c][0.7]{$\psi_1^-$}
\includegraphics[width=0.7\columnwidth]{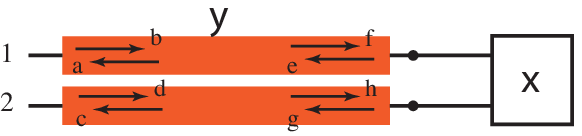}}
\subfigure[]{
\psfrag{a}[c][c][0.8]{$\theta_1, k_1$}
\psfrag{b}[c][c][0.8]{$\theta_2, k_2$}
\psfrag{c}[c][c][0.8]{$\theta_3, k_3$}
\psfrag{d}[c][c][0.8]{$\theta_i, k_i$}
\psfrag{e}[c][c][0.8]{$\theta_{N-1}, k_{N-1}$}
\psfrag{f}[c][c][0.8]{$\theta_N, k_N$}
\psfrag{p}[c][c][0.8]{$S_1$}
\psfrag{q}[c][c][0.8]{$S_2$}
\psfrag{r}[c][c][0.8]{$S_3$}
\psfrag{s}[c][c][0.8]{$S_i$}
\psfrag{t}[c][c][0.8]{$S_{N-1}$}
\psfrag{u}[c][c][0.8]{$S_N$}
\includegraphics[width=\columnwidth]{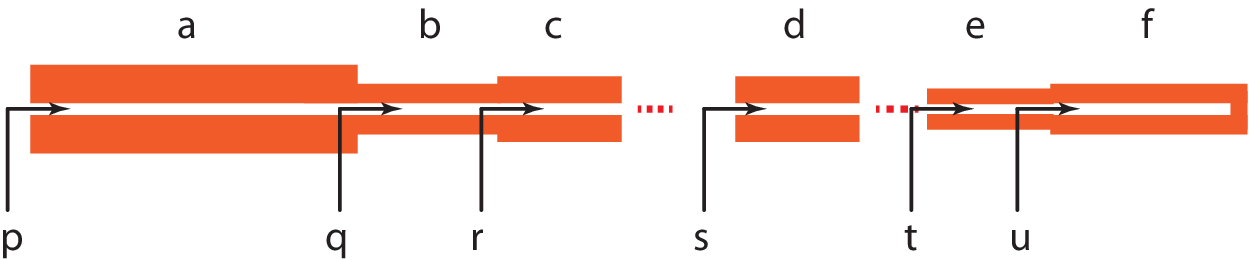}}
\caption{Coupled-line phaser configurations. a) C-section terminated with an arbitrary load of transfer function $S_0$. b) LC coupled-line phaser.} \label{Fig:CSec_Series}
\end{center}
\end{figure}

Consider an ideal lossless, perfectly matched and perfectly isolated TEM backward-wave coupled-line coupler, shown in Fig.~\ref{Fig:CSec_Series}(a). The 4-port scattering matrix of such a coupler is~\cite{Mongia_book_Couplers}

\begin{align}
\left[\begin{array}{c}
\psi^+_1\\
\psi^+_2\\
\psi^+_3\\
\psi^+_4\\
\end{array}\right]=
\left[\begin{array}{cccc}
0 & b(\theta) & a(\theta) & 0\\
b(\theta) & 0 & 0 & a(\theta)\\
a(\theta) & 0 & 0 & b(\theta)\\
0 & a(\theta) & b(\theta) & 0\\
\end{array}\right]
\left[\begin{array}{c}
\psi^-_1\\
\psi^-_1\\
\psi^-_1\\
\psi^-_1\\
\end{array}\right],\label{Eq:4portCC}
\end{align}

\noindent where

\begin{subequations}
\begin{equation}
b(\theta) = \frac{jk\sin\theta}{\sqrt{1-k^2}\cos\theta + j\sin\theta},
\end{equation}
\begin{equation}
a(\theta) = \frac{\sqrt{1-k^2}}{\sqrt{1-k^2}\cos\theta + j\sin\theta},
\end{equation}\label{Eq:s_1_s_2}
\end{subequations}

\noindent and $k$ is the coupling coefficient. The connection via a two-port of transfer function $S_0$ of ports 3 and 4 corresponds to the relationship

\begin{align}
\left[\begin{array}{c}
\psi^-_3\\
\psi^-_4\\
\end{array}\right]=
\left[\begin{array}{cc}
0 & S_0\\
S_0 & 0\\
\end{array}\right]
\left[\begin{array}{c}
\psi^+_3\\
\psi^+_4\\
\end{array}\right].\label{Eq:4portCC}
\end{align}

Subsequently prescribing $\psi^-_3 = S_0\psi^+_4$ and $\psi^-_4 = S_0\psi^+_3$ in \eqref{Eq:4portCC} transforms the four-port coupled-line coupler into a (two-port) C-section described by the following set of equations: $\psi^+_1 = b\psi^-_2 + aS_0\psi^+_4$, $\psi^+_2  = b\psi^-_1 + bS_0\psi^+_3$, $\psi^+_3 = a\psi^-_1/(1-bS_0)$ and $\psi^+_4 = a\psi^-_2/(1-bS_0)$. These relations lead to the two-port transfer function

\begin{equation}
S_{21}(\theta) = \frac{\psi^+_1}{\psi^-_2} = b + \frac{a^2S_0}{1-bS_0}.\label{Eq:S21_S0A}
\end{equation}

\noindent For $S_0=1$, this function may be explicitly written as

\begin{align}
S_{21}(\theta) &= \left(\frac{\sqrt{1+k}\cos\theta - j\sqrt{1-k}\sin\theta}{\sqrt{1+k}\cos\theta - j\sqrt{1-k}\sin\theta}\right)\notag\\
&= \left(\frac{\rho - j\tan\theta}{\rho + j\tan\theta}\right),\quad\text{where}\quad \rho = \sqrt{\frac{1 +k }{1 - k}}.
\end{align}

\subsection{Derivation of the  Longitudinally-cascaded (LC) Coupled-line Phaser Transfer Function [Eq.~(\ref{Eq:Series_TF})]}\label{Sec:Series_TF}

Consider the LC coupled-line sections of Fig.~\ref{Fig:CSec_Series}(b). The transfer function $S_N$ of the last coupled-section is given by

\begin{equation}
S_N(\theta)  = \left(\frac{\rho_N - j\tan\theta_N}{\rho_N + j\tan\theta_N}\right),
\end{equation}

\noindent where $\rho_N = \sqrt{1+k_N/1-k_N}$. Using (\ref{Eq:S21_S0}), with $S_0 = S_N$, the transfer function of the structure looking into the $(N-1)^\text{th}$ section is found as

\begin{equation}
S_{N-1}(\theta) = b_{N-1} + \frac{a_{N-1}^2S_N}{1-b_{N-1}S_N}.
\end{equation}

\noindent Similarly, the transfer function of the structure looking into the $(N-2)^\text{th}$ section can be written in terms of the $(N-1)^\text{th}$ section as

\begin{equation}
S_{N-2}(\theta) = b_{N-2} + \frac{a_{N-2}^2S_{N-1}}{1-b_{N-2}S_{N-1}}.
\end{equation}

\noindent Generalizing this procedure, the transfer function of the structure looking into the $i^\text{th}$ section can be iteratively expressed in terms of $(i+1)^\text{th}$-section as

\begin{subequations}
\begin{equation}
S_i(\theta)  = b_{i} + \frac{a_{i}^2S_{i+1}}{1- b_{i}S_{i+1}},
\end{equation}

with
\begin{equation}
b_{i}(\theta)  = \left(\frac{jk_i\sin\theta_i}{\sqrt{1-k_i^2}\cos\theta_i + j\sin\theta_i}\right)
\end{equation}
and
\begin{equation}
a_{i}(\theta)  = \left(\frac{\sqrt{1-k_i^2}}{\sqrt{1-k_i^2}\cos\theta_i + j\sin\theta_i}\right).
\end{equation}
\end{subequations}

\noindent Following this iterative procedure from end  to the input of the structure, the overall transfer function of an LC coupled-line phaser is found as

\begin{equation}
S_1(\theta) = b_{1} + \frac{a_{1}^2S_2(\theta) }{1- b_{1}S_2(\theta) }.
\end{equation}

\subsection{Proof of the Constant Area under the Group Delay Curves for TC-, LC and HC Coupled-Line Phasers}\label{Sec:ConstArea}

The proof the the area under the group delay curves for the TC-, LC and HC coupled-line phasers is constant may be conveniently in the low-pass Laplace domain. Consider first the transfer function of C- and D-sections, given by (\ref{Eq:Csection}) and (\ref{Eq:Dsection}), respectively, which become under the highpass-to-lowpass transformation $s = j\Omega = j\tan \theta$

\begin{subequations}
\begin{equation}
S_{21}^{\text{C-section}}(s) = \left(\frac{\rho - s}{\rho + s}\right) = \frac{H_1(s)}{H_1(-s)},\label{Eq:CS}
\end{equation}
\begin{equation}
S_{21}^{\text{D-section}}(s) = \left(\frac{\rho_i - s}{\rho_i + s}\right)\left(\frac{\rho_i^\ast - s}{\rho_i^\ast + s}\right)=\frac{H_2(s)}{H_2(-s)},\label{Eq:DS}
\end{equation}\label{Eq:CDS}
\end{subequations}

\noindent where $H_1(s)$ and $H_2(s)$ are first- and second-order Hurwitz polynomials, respectively, and $\rho_i$ is a function of $\rho_a$ and $\rho_b$, which are  defined in \eqref{Eq:rhoab}.

A general $N^\text{th}$-order all-pass phaser can be expressed in terms of an $N^\text{th}$-order Hurwitz polynomial, $H_N(s) = s^N + c_{N-1}S^{N-1} + \ldots + c_1s + c_0$, where $c_n \in\Re$, which can always by decomposed as a product of first and second order polynomials,

\begin{align}
S_{21}(s) &= \frac{H_N(s)}{H_N(-s)}= \prod_{m=1}^p \frac{\rho_m - s}{\rho_m+s}\prod_{n=1}^q \frac{\rho_n - s}{\rho_n+s}\frac{\rho_n^\ast - s}{\rho_n^\ast+s}.\label{Eq:NCDsection}
\end{align}

\noindent This means that any $N^\text{th}$-order all-pass phaser can be realized using $p$ C-sections and $q$ D-sections, such that $N = p+2q$. Consequently, the area under the $\tau-\theta$ curve of a $N^\text{th}$-order all-pass network can be deduced from those of its C- and D-section constituents.

The area under the group delay curve of an all-pass phaser in a half harmonic period (for symmetry) is

\begin{equation}
I = \int_0^{\pi/2} \tau(\theta)d\theta = \phi(0) - \phi(\pi/2),
\end{equation}

\noindent where the relation \mbox{$\tau(\theta) = -d\phi(\theta)/d\theta$} has been used. In the Laplace domain, following from $s = j\tan\theta$, the above expression becomes,

\begin{equation}
I =  \phi_s(0) - \phi_s(j\infty).
\end{equation}

\noindent For a single C-section, corresponding to a first-order polynomial in $s = \alpha + j\sigma$, the phase expression, using~\eqref{Eq:CS}, is

\begin{equation}
\phi_s(s) = \tan^{-1}\left(\frac{\sigma}{\alpha - \rho}\right) - \tan^{-1}\left(\frac{\sigma}{\alpha + \rho}\right).
\end{equation}

\noindent Since, subsequently,

\begin{align}
&\phi_s(0) = 0,\notag\\
&\phi_s(j\infty) = -2\left.\tan^{-1}\left(\frac{\sigma}{\rho}\right)\right|_{\sigma =\infty} = -\pi\notag,\\
\end{align}

\noindent we have $I =  \phi_s(0) - \phi_s(j\infty)  = \pi$. Thus, the area under the $\tau-\theta$ curve of a single C-section is $\pi$.

Similarly, from (\ref{Eq:DS}), the transmission phase of a single D-section

\begin{align}
\phi(s) &= \tan^{-1}\left(\frac{y-\sigma}{x-\alpha}\right) - \tan^{-1}\left(\frac{y+\sigma}{x-\alpha}\right)\\
& - \tan^{-1}\left(\frac{y+\sigma}{x+\alpha}\right) + \tan^{-1}\left(\frac{y-\sigma}{x+\alpha}\right),
\end{align}

\noindent using $\rho = x+ jy$. Since

\begin{align}
&\phi_s(0) = 0 \notag\\
&\phi_s(j\infty), = 2\left[\tan^{-1}\left(\frac{y-\infty}{x}\right) - \tan^{-1}\left(\frac{y+\infty}{x}\right)\right] = -2\pi\notag,\\
\end{align}

\noindent we have $I =  \phi_s(0) - \phi_s(j\infty)  = 2\pi$. Thus, the area under the $\tau-\theta$ curve of a single D-section is $2\pi$. 

From (\ref{Eq:NCDsection}), the area under the $\tau - \theta$ curve of a general phaser of the $N^\text{th}$ order is $p\pi + q(2\pi) = (p+2q)\pi = N\pi$, with ($p\pi$) contributed by the $p$ C-sections and ($2q\pi$) contributed by the $q$ D-sections. Finally, since the transfer function of the TC, LC and HC coupled-line phaser configurations can be all represented in terms of $H_N(s)$, the area under the $\tau-\theta$ curves is $N\pi$ in a given harmonic period for all three cases.

\section*{Acknowledgement}

This work was supported by NSERC Grant CRDPJ 402801-10 in partnership with Blackberry Inc and in part by HK ITP/026/11LP, HK GRF 711511, HK GRF 713011, HK GRF 712612, and NSFC 61271158.

\bibliographystyle{IEEEtran}
\bibliography{Gupta_CoupledLine_Phasers_TMTT_2013_Ref}

\begin{thebibliography}{10}
\providecommand{\url}[1]{#1}
\csname url@samestyle\endcsname
\providecommand{\newblock}{\relax}
\providecommand{\bibinfo}[2]{#2}
\providecommand{\BIBentrySTDinterwordspacing}{\spaceskip=0pt\relax}
\providecommand{\BIBentryALTinterwordstretchfactor}{4}
\providecommand{\BIBentryALTinterwordspacing}{\spaceskip=\fontdimen2\font plus
\BIBentryALTinterwordstretchfactor\fontdimen3\font minus
  \fontdimen4\font\relax}
\providecommand{\BIBforeignlanguage}[2]{{%
\expandafter\ifx\csname l@#1\endcsname\relax
\typeout{** WARNING: IEEEtran.bst: No hyphenation pattern has been}%
\typeout{** loaded for the language `#1'. Using the pattern for}%
\typeout{** the default language instead.}%
\else
\language=\csname l@#1\endcsname
\fi
#2}}
\providecommand{\BIBdecl}{\relax}
\BIBdecl

\bibitem{Caloz_MM_2012}
C.~Caloz, S.~Gupta, Q.~Zhang, and B.~Nikfal, ``Analog signal processing,''
  \emph{{Microw. Mag.}}, vol.~14, no.~6, pp. 87--103, Sept. 2013.

\bibitem{Caloz_PIEEE_10_2011}
C.~Caloz, ``Metamaterial dispersion engineering concepts and applications,''
  \emph{{Proc. IEEE}}, vol.~99, no.~10, pp. 1711--1719, Oct. 2011.

\bibitem{Lewis_SAW_OSP_2005}
M.~Lewis, ``{SAW} and optical signal processing,'' in \emph{Proc. {IEEE}
  Ultrason. Symp.}, vol.~24, Sept. 2005, pp. 800--809.

\bibitem{Abielmona_TMTT_11_2009}
S.~Abielmona, S.~Gupta, and C.~Caloz, ``Compressive receiver using a
  {CRLH}-based dispersive delay line for analog signal processing,''
  \emph{{IEEE Trans. Microw. Theory Tech.}}, vol.~57, no.~11, pp. 2617--2626,
  Nov. 2009.

\bibitem{Nguyen_MWCL_08_2008}
H.~V. Nguyen and C.~Caloz, ``Composite right/left-handed delay line pulse
  position modulation transmitter,'' \emph{{IEEE Microw. Wireless Compon.
  Lett.}}, vol.~18, no.~5, pp. 527--529, Aug. 2008.

\bibitem{Nikfal_TMTT_06_2011}
B.~Nikfal, S.~Gupta, and C.~Caloz, ``Increased group delay slope loop system
  for enhanced-resolution analog signal processing,'' \emph{{IEEE Trans.
  Microw. Theory Tech.}}, vol.~59, no.~6, pp. 1622--1628, Jun. 2011.

\bibitem{Nikfal_MWCL_11_2012}
B.~Nikfal, D.~Badiere, M.~Repeta, B.~Deforge, S.~Gupta, and C.~Caloz,
  ``Distortion-less real-time spectrum sniffing based on a stepped group-delay
  phaser,'' \emph{{IEEE Microw. Wireless Compon. Lett.}}, vol.~22, no.~11, pp.
  601--603, Oct. 2012.

\bibitem{Gupta_TMTT_04_2009}
S.~Gupta, S.~Abielmona, and C.~Caloz, ``Microwave analog real-time spectrum
  analyzer ({RTSA}) based on the spatial-spectral decomposition property of
  leaky-wave structures,'' \emph{{IEEE Trans. Microw. Theory Tech.}}, vol.~59,
  no.~12, pp. 2989--2999, Dec. 2009.

\bibitem{Schwartz_MWCL_04_2006}
J.~D. Schwartz, J.~Aza{\~{n}}a, and D.~Plant, ``Experimental demonstration of
  real-time spectrum analysis using dispersive microstrip,'' \emph{{IEEE
  Microw. Wireless Compon. Lett.}}, vol.~16, no.~4, pp. 215--217, Apr. 2006.

\bibitem{Laso_TMTT_03_2003}
M.~A.~G. Laso, T.~Lopetegi, M.~J. Erro, D.~Benito, M.~J. Garde, M.~A. Muriel,
  M.~Sorolla, and M.~Guglielmi, ``Real-time spectrum analysis in microstrip
  technology,'' \emph{{IEEE Trans. Microw. Theory Tech.}}, vol.~51, no.~3, pp.
  705--717, Mar. 2003.

\bibitem{Xiang_TMTT_11_2012}
B.~Xiang, A.~Kopa, F.~Zhongtao, and A.~B. Apsel, ``Theoretical analysis and
  practical considerations for the integrated time-stretching system using
  dispersive delay line ({DDL}),'' \emph{{IEEE Trans. Microw. Theory Tech.}},
  vol.~60, no.~11, pp. 3449--3457, Nov. 2012.

\bibitem{Schwartz_MWCL_01_2008}
J.~D. Schwartz, I.~Arnedo, M.~A.~G. Laso, T.~Lopetegi, J.~Aza{\~{n}}a, and
  D.~Plant, ``An electronic {UWB} continuously tunable time-delay system with
  nanosecond delays,'' \emph{{IEEE Microw. Wireless Compon. Lett.}}, vol.~18,
  no.~2, pp. 103--105, Jan. 2008.

\bibitem{Gupta_AWPL_11_2011}
S.~Gupta, B.~Nikfal, and C.~Caloz, ``Chipless {RFID} system based on group
  delay engineered dispersive delay structures,'' \emph{{IEEE Antennas Wirel.
  Propagat. Lett.}}, vol.~10, pp. 1366--1368, Dec. 2011.

\bibitem{Laso_MWCL_12_2001}
M.~A.~G. Laso, T.~Lopetegi, M.~J. Erro, D.~Benito, M.~J. Garde, M.~A. Muriel,
  M.~Sorolla, and M.~Guglielmi, ``Chirped delay lines in microstrip
  technology,'' \emph{{IEEE Microw. Wireless Compon. Lett.}}, vol.~11, no.~12,
  pp. 486--488, Dec. 2001.

\bibitem{Coulombe_TMTT_08_2009}
M.~Coulombe and C.~Caloz, ``Reflection-type artificial dielectric substrate
  microstrip dispersive delay line ({DDL}) for analog signal processing,''
  \emph{{IEEE Trans. Microw. Theory Tech.}}, vol.~57, no.~7, pp. 1714--1723,
  Jul. 2009.

\bibitem{Zhang_TMTT_08_2012}
Q.~Zhang, S.~Gupta, and C.~Caloz, ``Synthesis of narrow-band reflection-type
  phaser with arbitrary prescribed group delay,'' \emph{{IEEE Trans. Microw.
  Theory Tech.}}, vol.~60, no.~8, pp. 2394--2402, Aug. 2012.

\bibitem{Zhang_EL_14_2013}
Q.~Zhang and C.~Caloz, ``Comparison of transmission and reßection allpass
  phasers for analogue signal processing,'' \emph{{Electron. Lett.}}, vol.~49,
  no.~14, Jul. 2013.

\bibitem{Campbell_SAW_1989}
C.~K. Campbell, \emph{Surface Acoustic Wave Devices and Their Signal Processing
  Applications}.\hskip 1em plus 0.5em minus 0.4em\relax Academic Press, 1989.

\bibitem{Ishak_PIEE_02_1988}
W.~S. Ishak, ``Magnetostatic wave technology: a review,'' \emph{{Proc. IEEE}},
  vol.~76, no.~2, pp. 171--187, Feb. 1988.

\bibitem{Zhang_TMTT_3_2013}
Q.~Zhang, D.~L. Sounas, and C.~Caloz, ``Synthesis of cross-coupled
  reduced-order phasers with arbitrary group delay and controlled magnitude,''
  \emph{{IEEE Trans. Microw. Theory Tech.}}, vol.~61, no.~3, pp. 1043--1052,
  Mar. 2013.

\bibitem{Atia_TMTT_08_2002}
H.-T. Hsu, H.-W. Yao, K.~A. Zaki, and A.~E. Atia, ``Synthesis of
  coupled-resonators group-delay equalizers,'' \emph{{IEEE Trans. Microw.
  Theory Tech.}}, vol.~50, no.~8, pp. 1960--1968, Aug. 2002.

\bibitem{Cristal_TMTT_06_1966}
E.~G. Cristal, ``Analysis and exact synthesis of cascaded commensurate
  transmission-line {C}-section all-pass networks,'' \emph{{IEEE Trans. Microw.
  Theory Tech.}}, vol.~14, no.~6, pp. 285--291, Jun. 1966.

\bibitem{Gupta_TMTT_09_2010}
S.~Gupta, A.~Parsa, E.~Perret, R.~V. Snyder, R.~J. Wenzel, and C.~Caloz,
  ``Group delay engineered non-commensurate transmission line all-passnetwork
  for analog signal processing,'' \emph{{IEEE Trans. Microw. Theory Tech.}},
  vol.~58, no.~9, pp. 2392--2407, Sept. 2010.

\bibitem{Horii_MWCL_01_2012}
Y.~Horii, S.~Gupta, B.~Nikfal, and C.~Caloz, ``Multilayer broadside-coupled
  dispersive delay structures for analog signal processing,'' \emph{{IEEE
  Microw. Wireless Compon. Lett.}}, vol.~22, no.~1, pp. 1--3, Jan. 2012.

\bibitem{Gupta_IJCTA_2013}
S.~Gupta, D.~L. Sounas, Q.~Zhang, and C.~Caloz, ``All-pass dispersion synthesis
  using microwave {C-sections},'' \emph{{Int. J. Circ. Theory Appl.}}, May
  2013.

\bibitem{Zhang_IJRMCAE_2013}
Q.~Zhang, S.~Gupta, and C.~Caloz, ``Synthesis of broadband phasers formed by
  commensurate {C- and D-sections},'' \emph{{Int. J. RF Microw. Comput. Aided
  Eng.}}, Aug. 2013.

\bibitem{Gupta_Folded_ICEAA_2013}
S.~Gupta, L.~J. Jiang, and C.~Caloz, ``Enhanced-resolution folded {C}-section
  phaser,'' in \emph{{Int. Conf. on Electromagnetics Advanced Applications
  (ICEAA)}}, Sept. 2013, pp. 771--773.

\bibitem{Paradis_EL_00_2013}
T.~Paradis, S.~Gupta, Q.~Zhang, L.~J. Jiang, and C.~Caloz, ``Hybrid-cascade
  coupled-line phasers for high-resolution radio-analog signal processing,''
  \emph{{Electron. Lett.}}, 2013, submitted.

\bibitem{Steenart_TMTT_01_1963}
W.~J.~D. Steenaart, ``The synthesis of coupled transmission line all-pass
  networks in cascades of 1 to $n$,'' \emph{{IEEE Trans. Microw. Theory
  Tech.}}, vol.~11, no.~1, pp. 23--29, Jan. 1963.

\bibitem{Cristal_TMTT_01_1969}
E.~G. Cristal, ``Theory and design of transmission line all-pass equalizers,''
  \emph{{IEEE Trans. Microw. Theory Tech.}}, vol.~17, no.~1, pp. 28--38, Jan.
  1969.

\bibitem{Zhang_APM_2_2013}
Q.~Zhang, D.~L. Sounas, S.~Gupta, and C.~Caloz, ``Wave interference explanation
  of group delay dispersion in resonators,'' \emph{{IEEE Antennas Propagat.
  Mag.}}, vol.~55, no.~2, pp. 212--227, May 2013.

\bibitem{Gupta_TMTT_12_2012}
S.~Gupta, D.~L. Sounas, H.~V. Nguyen, Q.~Zhang, and C.~Caloz, ``{CRLH-CRLH}
  {C-section} dispersive delay structures with enhanced group delay swing for
  higher analog signal processing resolution,'' \emph{{IEEE Trans. Microw.
  Theory Tech.}}, vol.~60, no.~21, pp. 3939--3949, Dec. 2012.

\bibitem{Mongia_book_Couplers}
R.~K. Mongia, I.~J. Bahl, P.~Bhartia, and J.~Hong, \emph{RF and Microwave
  Coupled-Line Circuit}.\hskip 1em plus 0.5em minus 0.4em\relax Artech House
  Publishers, 2nd Ed., May. 2007.

\end{thebibliography}

\end{document}